\documentclass{article}

\usepackage{microtype}
\usepackage{graphicx}
\usepackage{subfigure}
\usepackage{booktabs} 

\usepackage{hyperref}

\usepackage[accepted]{arxiv2025}

\usepackage{amsmath}
\usepackage{amssymb}
\usepackage{mathtools}
\usepackage{amsthm}

\usepackage{xcolor}         
\usepackage{graphicx}
\usepackage{algpseudocode}
\usepackage{algorithm}
\usepackage{indentfirst}
\usepackage{extarrows}

\usepackage{tikz}
\newcommand*\circled[1]{\tikz[baseline=(char.base)]{
    \node[shape=circle,draw,inner sep=1pt] (char) {#1};}}

\usepackage[capitalize,noabbrev]{cleveref}

\theoremstyle{plain}

\theoremstyle{definition}

\theoremstyle{remark}

\usepackage[textsize=tiny]{todonotes}

\arxivtitlerunning{Nanoscaling Floating-Point (NxFP): Improving Microscaling Standard using 3 Techniques}

\begin{document}

\twocolumn[
\arxivtitle{Nanoscaling Floating-Point (NxFP): NanoMantissa, Adaptive Microexponents, and Code Recycling for Direct-Cast Compression of Large Language Models}



\arxivsetsymbol{equal}{*}

\begin{arxivauthorlist}

\textbf{Yun-Chen Lo \quad Gu-Yeon Wei \quad David Brooks }\\
Harvard University
\end{arxivauthorlist}

\arxivkeywords{Machine Learning, arxiv}

\vskip 0.3in
]

\makeatletter{\renewcommand*{\@makefnmark}{}
\footnotetext{Preprint. }\makeatother}

\begin{abstract}
As cutting-edge large language models (LLMs) continue to transform various industries, their fast-growing model size and sequence length have led to memory traffic and capacity challenges. 
Recently, AMD, Arm, Intel, Meta, Microsoft, NVIDIA, and Qualcomm have proposed a Microscaling standard (Mx), which augments block floating-point with microexponents to achieve promising perplexity-to-footprint trade-offs.
However, the Microscaling suffers from significant perplexity degradation on modern LLMs with less than six bits. 

This paper profiles modern LLMs and identifies three main challenges of low-bit Microscaling format, i.e., inaccurate tracking of outliers, vacant quantization levels, and wasted binary code.
In response, \textbf{Nanoscaling (NxFP)} proposes three techniques, i.e., \emph{NanoMantissa}, \emph{Adaptive Microexponent}, and \emph{Code Recycling} to enable better accuracy and smaller memory footprint than state-of-the-art MxFP. 
Experimental results on direct-cast inference across various modern LLMs demonstrate that our proposed methods outperform state-of-the-art MxFP by up to 0.64 in perplexity and by up to 30\% in accuracy on MMLU benchmarks. 
Furthermore, NxFP reduces memory footprint by up to 16\% while achieving comparable perplexity as MxFP. 
\end{abstract}

\section{Introduction}

Large language models (LLMs) have successfully enabled more and more ground-breaking applications, e.g., general assistant~\cite{2023_arxiv_llama, 2020_arxiv_gpt3}, automatic code generation~\cite{2023_arxiv_starcoder}, and humanoid robots~\cite{2024_chi_robotllm}. Following the promising neural scaling law~\cite{2022_nips_neural_scaling,2020_arxiv_neural_scaling_2}, next-generation LLMs continue to scale up in parameter counts, training computation, training data, and sequence length. However, the growth rate of model size (410$\times$/2years) dramatically outweighs the scaling speed of DRAM traffic (1.4$\times$/2years) and capacity (2$\times$/2years), leading to a "memory wall" challenge~\cite{2024_ieeemicro_memorywall}. 

\begin{figure}[t]
    \centering
    \includegraphics[width=1.0\linewidth]{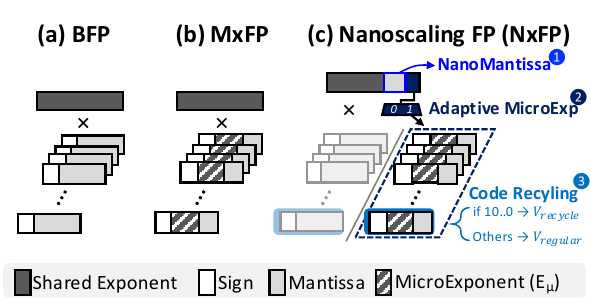}
    \caption{(a) Block FP, (b) Microscaling FP, and (c) our Nanoscaling FP (NxFP). NxFP proposes NanoMantissa, Adaptive MicroExponent, and  Code Recycling to outperform the MxFP standard.
    }
    \label{figure:proposed}
\end{figure}

To reduce the memory footprint, researchers have extensively explored various number systems, including integer~\cite{2023_iclr_gptq, 2024_mlsys_awq, 2024_arxiv_gptvq}, floating-point~\cite{2020_dac_adaptivfloat, 2023_tecs_sgfloat, 2023_cal_lv, 2024_arxiv_fp6llm}, and block floating-point (BFP)~\cite{2020_nips_msfp, 2017_nips_flexpoint, 2023_iclr_bsfp, 2023_micro_bucketgetter, 2024_icml_bie} (Figure~\ref{figure:proposed}(a)). Recently, AMD, Arm, Intel, Meta, Microsoft, NVIDIA, and Qualcomm further proposed \textbf{Microscaling (Mx)} format as a new standard to achieve state-of-the-art Pareto frontier in quantization error and memory footprint~\cite{2023_isca_microexponets, 2023_ocp_mx, 2023_arxiv_microscaling}. The Microscaling standard serves as a direct-cast compression method, which can used to quantize weights, KV cache, or even gradient to reduce the footprint.
As shown in Figure~\ref{figure:proposed}(b), the key of Microscaling format is to \emph{augment block floating-point format with microexponents ($E_{\mu}$)~\cite{2023_isca_microexponets} as second-level, per-element scaling factors}. 

While the Microscaling standard has been quickly adopted in various commercial products~\cite{2024_qualcomm_mx, 2024_intel_mx, 2024_nvidia_blackwell}, we identify that the Microscaling format significantly degrades the quality of generative inference when the bitwidth is lower than 6. For example, quantizing Llama using MxFP4 degrades the perplexity from 9.488 (FP16) to 27.201 (MxFP4), as presented in table 6 of the original paper~\cite{2023_arxiv_microscaling}. 
Therefore, \emph{how to rescue perplexity for generative language models using sub-6-bit Microscaling standard} remains a challenging question.

This paper aims to enhance the quality of sub-6-bit Microscaling format by introducing a next-generation custom format in the Microscaling-family for LLMs.
Based on profiling modern LLMs, we identify three main challenges on low-bit MxFP: 
(a) Unable to accurately represent significant values larger than the most considerable quantization value. 
(b) Vacant quantization level due to the mismatch between distribution and format. 
(c) Wasted code because of $\pm$0 in sign-magnitude.
In response, we propose \textbf{Nanoscaling Floating-Point (NxFP)}, a novel datatype that addresses these challenges and outperforms the existing Microscaling standard.

Figure~\ref{figure:proposed}(c) presents our proposed NxFP, which obtains three enabling techniques, i.e., NanoMantissa, Adaptive Microexponent, and Code Recycling. 
We will show that NxFP is superior to MxFP at approximating real weights on LLMs (Figure~\ref{figure:qerr}). 
In addition, we show that NxFP has an efficient on-the-fly dequantization kernel, quantization flow, and structural memory layout for frictionless deployment.

Our contributions are summarized as follows:
\begin{itemize}

\item We profile modern LLMs and identify three fundamental challenges of the low-bit MxFP, i.e., inaccurate tracking of outliers, vacant quantization levels, and wasted binary code.

\item We propose \emph{NanoMantissa}, which allocates a 2-bit mantissa in the shared scaling factor to provide some precision in approximating large value in each block. 

\item We propose \emph{Adaptive Microexponent} to provide each vector the flexibility to select its most suitable format (MxFP/BFP) that can best fit the distribution.

\item We propose \emph{Code recycling} to remap the wasted code (i.e., -0 in sign-magnitude) to be half of the smallest quantization level and improve the quantization error.

\item We present a quantization flow and an on-the-fly dequantization kernel for efficient NxFP deployment on off-the-shelf hardware.

\item We perform extensive evaluations on various LLMs and demonstrate that NxFP successfully outperforms MxFP in memory footprint, quantization error, perplexity, and reasoning accuracy. 

\end{itemize}

\begin{figure}[t]
    \centering
    \includegraphics[width=0.91\linewidth]{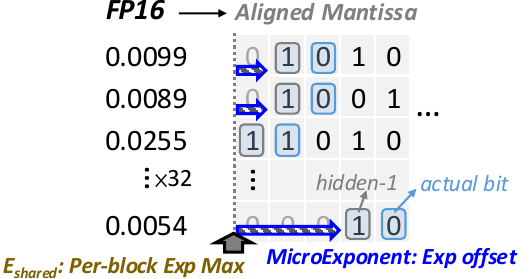}
    \caption{
    Visualizing quantization of a real FP16 vector using MxFP. The shared exponent tracks the largest exponent value in each block, and the microexponents track the exponent offset relative the shared exponent.
    }
    \label{figure:mx_format}
\end{figure}

\section{Understanding Microscaling (Mx) Standard}

Open Compute Project (OCP) has defined four standards in the Microscaling family, i.e., MxINT8, MxFP8 (E5M2/E4M3), MxFP6 (E3M2/E2M3), and MxFP4 (E2M1)~\cite{2023_ocp_mx}, where the suffix denotes the element format and the bitwidth. In addition, MxFP8 and MxFP6 have defined multiple optimized configurations to optimize for different scenarios. For example, MxFP8 with 4-bit exponent and 3-bit mantissa (E4M3) prioritizes precision and is more suitable for inference. 

The high-level ideas for Microscaling formats are summarized as follows. 
First, the MxINT8 is fundamentally BFP which replaces sign magnitude with two's complement.
Second, MxFP augments conventional BFP with microexponents. Hence, MxFP incorporates two exponent fields per vector: one shared exponent and multiple microexponents. The shared exponent is responsible for vector-wise scaling, and the microexponents serve as element-wise scaling.

Figure~\ref{figure:mx_format} visualizes quantizing a FP16 vector to a MxFP4 vector (E2M1).
First, given an FP16 vector, we obtain the mantissa binary and align each mantissa with the most significant exponent value.
Second, we configure the shared exponent as the largest exponent value in the vector (vertical arrow).
Third, the microexponents serve as exponent offsets to track each element's leading one (horizontal arrow).
Finally, the mantissa slices the trailing bit (blue rectangle) while assuming the leading one exists (grey rectangle). 
In sum, the MxFP format balances vector-wise dynamic range, element-wise dynamic range, and element-wise precision.

\begin{figure}[t]
    \centering
    \includegraphics[width=0.95\linewidth]{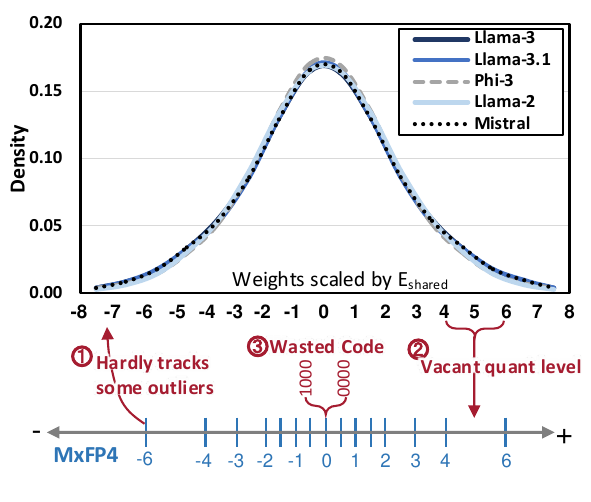}
    \caption{Profiling weights scaled by $E_{shared}$ (block size=32) on five modern LLMs. We show three challenges of MxFP4 format, including vacant quantization level, wasted code, and inefficient tracking of outliers.
    }
    \label{figure:challenge}
\end{figure}

\section{Challenges of Low-bit MxFP}

There are three challenges of low-bit MxFP, i.e., inaccurate tracking of largest values, vacant quantization level, and wasted code. To understand these challenges, we profile the distribution of weight values after its being scaled by $E_{shared}$ for each 32-wide weight vector on five modern LLMs (Llama3, Llama3.1, Phi3, Llama2, and Mistral). 

As shown in Figure~\ref{figure:challenge}, the scaled weights have a normal distribution ranging from -8$\sim$+8 that has a relatively large standard deviation. The element format, FP4, quantizes these full-precision values into quantization levels ranging from -6 to +6. 
The first challenge is that the shared exponent cannot efficiently scale the weight value to capture the largest value. For example, the maximum quantization level of MxFP4 is 6, which can hardly quantify weights with values 7$\sim$8.
Second, since MxFP4 allocates more quantization levels near zero, there is a vacant quantization levels in between 4 to 6. Values in this range can not be quantized efficiently, leading to significant quantization errors.
Third, MxFP4 adopts sign-magnitude and wastes two binary codes to represent -0 and +0, where its overheads grow as we scale down the bitwidth.

\begin{figure}[t]
    \centering
    \includegraphics[width=0.95\linewidth]{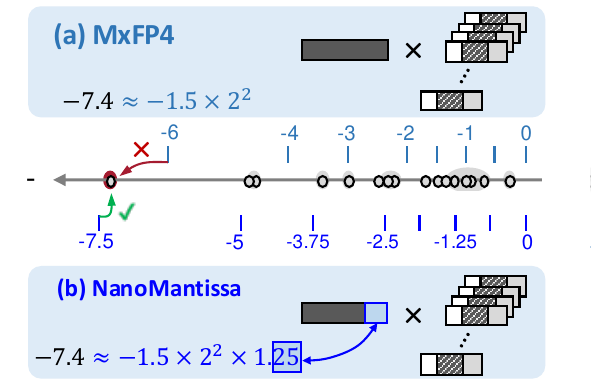}
    \caption{
    (a) MxFP4 and
    (b) MxFP4 with NanoMantissa. The proposed NanoMantissa enable MxFP4 with greater precision so that it can track the largest value more accurately.
    }
    \label{figure:nanomant}
\end{figure}

\section{Nanoscaling Floating-Point (NxFP)} \label{sec:Nx}

We present three enabling techniques of the NxFP format, i.e., NanoMantissa, Adaptive Microexponent, and Code Recycling.
We will provide a comprehensive overview of these techniques, offering a high-level understanding, and then delve into specific examples to illustrate them.

\subsection{NanoMantissa (NM)} \label{subsec:nanomant}

To help low-bit MxFP accurately track the largest value in each block, we propose incorporating a 2-bit mantissa field in the shared scaling factor. Conceptually, the 2-bit shared NanoMantissa helps low-bit MxFP formats boost their precision and track the full-precision values more accurately.

Figure~\ref{figure:nanomant} presents an example of quantizing Llama3 vector using (a) pure MxFP4 and (b) MxFP4 with NanoMantissa. While the most considerable scaled weight value is -7.4, MxFP can only approximate it using -6 ($-1.5\times2^2$). 
If we enable MxFP4 with a NanoMantissa, we can configure the NanoMantissa to scale the entire vector by 1.25, which best approximates -7.4 using -7.5. As a result, the L1 quantization error is significantly reduced from 1.4 to 0.1. We will show in Figure~\ref{figure:qerr} that NanoMantissa can reduce the MSE by 23\% when quantizing weights on modern LLMs. 

To set up the NanoMantissa value, the user can set the NanoMantissa as zero to nullify it. Otherwise, our quantization algorithm will round the most considerable value to get the NanoMantisa value and confirm that it improves the MSE (Algorithm~\ref{algo:quant_flow}).

\begin{figure}[t]
    \centering
    \includegraphics[width=0.95\linewidth]{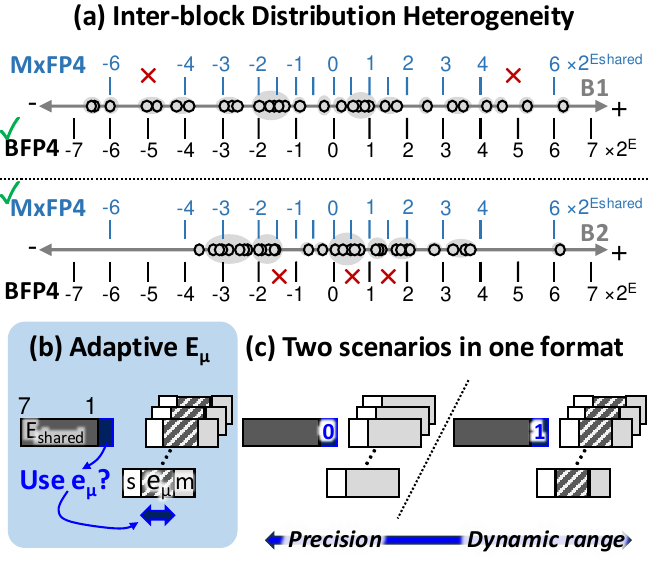}
    \caption{
    (a) Inter-block distribution heterogeneity, meaning that different vectors have distinct distribution, motivating each block to have its optimized format, e.g., MxFP4 or BFP4.
    (b) We propose to use an index bit to indicate whether using MxFP or BFP for a target block.
    (c) Logically, we fuse two formats into one, which adapts to different scenarios.
    }
    \label{figure:ada_uexp}
\end{figure}

\subsection{Adaptive Microexponent (AM)} \label{subsec:adaexp}

To mitigate the challenge of vacant quantization level, we propose Adaptive Microexponent, which only allocates microexponent fields for vectors with scattered distribution. In other word, we successfully mitigate the vacant quantization level issue by choosing a suitable format for each block.

Figure~\ref{figure:ada_uexp}(a) presents that different blocks have distinct distributions, where the first block is more clustered in values than the second block. We quantize each block using MxFP4 and BFP4 to showcase that vacant quantization level issue appeals when there is a mismatch between the low-bit format and the per-block distribution. 
When quantizes value-clustered block 1 (B1), BFP4 outperforms MxFP4 because the latter suffer from vacant quantization level around 5 (as denoted by the red cross in MxFP4).
When quantizing value-scattered block 2 (B2), MxFP4 outperforms BFP4 because BFP4 suffers from a more significant quantization error around 0 (as denoted by the red cross in BFP4). 

Figure~\ref{figure:ada_uexp}(b) presents the adaptive microexponent scheme, which augments each Microscaling vector with an index bit to infer whether a block uses a microexponent or not. Doing this gives each block the flexibility to choose its best format.

Figure~\ref{figure:ada_uexp}(c) illustrates two possible formats. When a given vector is clustered in value, we can set the index bit as 0 to choose BFP format, which allocates all the bits to be mantissas and optimizes precision. 
Conversely, when a vector contains outliers, setting the index bit as 1 configures the block as MxFP, ensuring a great dynamic range and more quantization levels around zero.

\begin{figure}[t]
    \centering
    \includegraphics[width=0.92\linewidth]{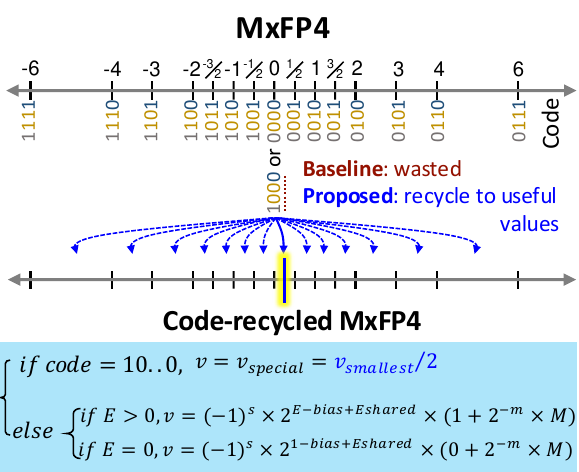}
    \caption{
    The flow to apply code recycling on MxFP4. Considering the perplexity improvement and the implementation overheads, we remaps -0 (code: 1000) to $\frac{1}{2}V_{smallest}$.
    }
    \label{figure:code_recycle}
\end{figure}

\subsection{Code Recycling (CR)} \label{subsec:coderecycle}
We propose code recycling to mitigate the wasted code issue in sign-magnitude format. 
Our approach involves remapping the binary code representing -0 in sign-magnitude numbers to a useful value, which improves the quantization error.

The blue arrows in Figure~\ref{figure:code_recycle} present the process of sweeping for different remapped values. 
A good remapped value should simultaneously improve the perplexity and have low implementation overheads. We set the remapped value to be $\frac{1}{2}\times V_{smallest}$ based on our empirical profiling of three representative LLMs (Llama2, Llama3, and Llama3.1). During the dequantization, we can right-shift the smallest number by one bit to obtain the remapped value during decoding.

\section{Quantization Algorithm} \label{sec:quant}

Similar to MxFP standard, NxFP format is not limited to just one quantization algorithm. For example, we present in below an MSE-based quantization algorithms, which maximizes the accuracy.

\textbf{MSE-based quantization algorithm:} 
Algorithm~\ref{algo:quant_flow} presents the process of quantizing a full-precision vector into a low-bit Nanoscaling vector.
First, we find the largest absolute value $V_{max}$ from the full-precision vector $V_{fp}$. 
Second, we find the largest exponent $E_{max}$.
Third, we normalize the largest absolute value and round to get a 2-bit NanoMantissa for scaling.
Fourth, we scale the original vector and quantize it into MxFP and BFP formats. 
Then, we evaluate the quantization error (MSE) to determine which format is better and set the format indicator bit $fmt$.
The above quantization is repeated with NanoMantissa set to zero to get the MSE-optimized binary. Please note that the quantization function contains code recycling, which maps -0 to be half of the smallest quantization level.

\begin{algorithm} [ht!]
\caption{MSE-based Quantization Algorithm}\label{alg:cap}
\textbf{Input:} Full-precision vector $V_{fp}$ \\
\textbf{Output:} NxFP vector with scaling ($1.M_{nano}\times E_{shared}$), format $fmt$, and element $V_{nx}$
\begin{algorithmic}
\Require Element bits $B$ and Microexponent bits $B_{E\mu}$
\State $V_{max} \gets Max(Abs(V_{fp}))$
\State $E_{max} \gets Log2(V_{max})$
\State $M_{nano} \gets Round_{2b}((V_{max}\gg E_{max})\ll 2$)
\State $MSE \gets MAX$
\For {$m \in $\{$M_{nano}, 0\} $}
\State $V_{scaled} = V_{fp}\gg E_{max}/(1.m)$ 
\State $V_{mxfp}, E_{shared} \gets Round_{mxfp}(V_{scaled},~B)$
\State $V_{bfp}, E_{shared} \gets Round_{bfp}(V_{scaled},~B)$ 
\State $MSE_{mxfp} \gets MSE(V_{mxfp},V_{scaled})$ 
\State $MSE_{bfp} \gets MSE(V_{bfp},V_{scaled})$ 

\If{$MSE_{mxfp}<MSE_{bfp}$}.  
    \If{$MSE_{mxfp}<MSE$}.
        \State $V_{nx} = V_{mxfp}$
        \State $fmt = 1$
        \State $MSE=MSE_{mxfp}$
    \EndIf
\Else
    \If{$MSE_{bfp}<MSE$}.
        \State $V_{nx} = V_{bfp}$
        \State $fmt = 0$
        \State $MSE=MSE_{bfp}$
    \EndIf
\EndIf
\EndFor
\end{algorithmic}
\label{algo:quant_flow}
\end{algorithm}

\begin{figure*}[t]
    \centering
    \includegraphics[width=0.82\linewidth]{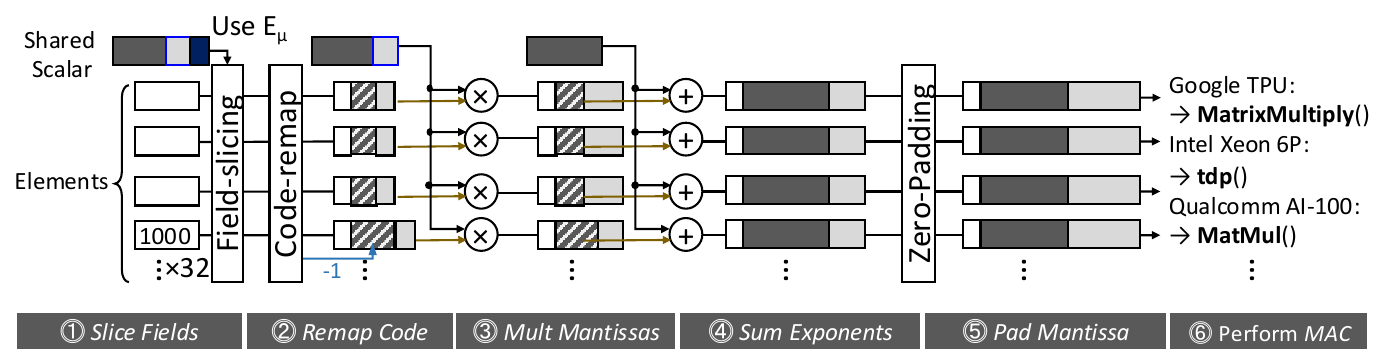}
    \caption{Six steps to decompress Nanoscaling format and run on off-the-shelf hardware. The only difference of NxFP, compared to Mx, is to slice fields, remap wasted code, and multiply NanoMantissa with per-element mantissa. }
    \label{figure:decompress}
\end{figure*} 

\section{On-the-fly Nanoscaling Dequantization  on Off-the-shelf Hardware} \label{sec:otf}

On-the-fly dequantization is an intriguing technique for deploying Microscaling format on off-the-shelf AI accelerators. 
This deployment method reduces memory footprint while maintaining great hardware compatibility.
Specifically, we allow data to be quantized into Mx format for storage in off-chip memory and then decompressed for execution on FP16/BF16 cores. For instance, Qualcomm's AI-100 accelerator supports MxFP6 on-the-fly dequantization to achieve approximately $2\times$ speedup over FP16~\cite{2024_qualcomm_mx}. Intel's Xeon processors also support MxFP4 on-the-fly-dequantization to improve the token generation speed~\cite{2024_intel_mx}.

Similarly, NxFP can also employ an on-the-fly dequantization method to run on off-the-shelf hardware. 
Figure~\ref{figure:decompress} presents the dequantization flow of the Nanoscaling format, in which we discuss the detailed operations below:

\textbf{\circled{1} Slice the fields:} \thickspace \thickspace
The format index bit ($fmt$) is used to identify the format of a given block and slice out the corresponding fields.
\textbf{\circled{2} Decode the wasted code.} \thickspace \thickspace
If the element code is 10...0, we remap it to half of the smallest value by right-shifting the smallest quantization value by one bit.
\textbf{\circled{3} Multiply the mantissas.} \thickspace \thickspace
We multiply the NanoMantissa with the element-wise mantissa to get the expanded mantissa field.
\textbf{\circled{4} Sum exponents.} \thickspace \thickspace
We add the microexponents with the shared exponent to form the expanded exponent field, as multiplying two power-of-two numbers equates to summing the exponents.
\textbf{\circled{5} Pad the mantissa.} \thickspace \thickspace
For conversion to BF16 (1s8e7m) or FP16, the mantissa, and exponent are properly padded with zeros as LSBs.
\textbf{\circled{6} Perform MAC.} \thickspace \thickspace
The final step is to initiate computation using appropriate instructions for different platforms, such as calling MatrixMultiply() on Google TPU or tdp() on the Intel Xeon 6 processor.

\section{Experiments} \label{sec:exp}

\subsection{Settings} \label{sec:setup}

\textbf{Quantization.} \thickspace \thickspace 
This work focuses on direct-cast inference, which directly reflects the efficiency of different number systems. Please note that our proposed number system is orthogonal and can be integrated with many post-training quantization techniques. Further, direct-cast quantization minimizes the deployment friction and avoids the risk of overfitting to the calibration set. 

We present the results of 1) quantizing only the weights, and 2) quantizing the weights and KV cache because they dominate the memory footprint~\cite{2024_arxiv_kvquant}.
We used a block size of 32 unless explicitly specified, where 32 is officially defined in Microscaling standard~\cite{2023_ocp_mx}.

\textbf{Models.} \thickspace \thickspace 
We benchmark our method on numerous mainstream large language models, including Llama3 (8B)~\cite{2024_web_llama3}, Llama3.1 (8B)~\cite{2024_arxiv_llama3p1}, Phi3 (4B)~\cite{2024_arxiv_phi3}, Gemma2~\cite{2024_arxiv_gemma2} 
, Llama2 (7B)~\cite{2023_arxiv_llama2}, Llama2 (13B)~\cite{2023_arxiv_llama2}, and Mistral (7B)~\cite{2023_arxiv_mistral}. These representative open-source models are trained using different datasets, parameter counts, number of operations, and model architectures. 

\textbf{Baselines.} \thickspace \thickspace 
Our primary baseline is Microscaling~\cite{2023_ocp_mx} because it can achieve state-of-the-art trade-offs and is utilized by commercial products. We evaluate different microexponent configurations and choose to report the best perplexity.
In addition, we also compare with block floating point (BFP~\cite{2020_nips_msfp}, or MxINT~\cite{2023_ocp_mx}). 

\begin{figure}[t]
    \centering
    \includegraphics[width=0.95\linewidth]{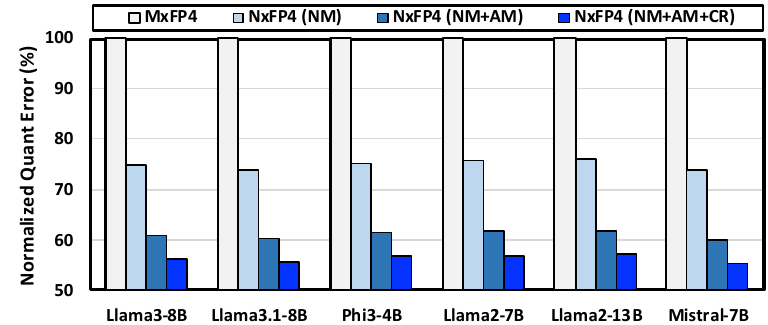}
    \caption{NxFP4 consistently reduces 10\%$\sim$14\% quantization error over MxFP4.}
    \label{figure:qerr}
\end{figure}

\textbf{Evaluations.} \thickspace \thickspace 
We evaluate the quantized LLMs on language generative performance using perplexity on Wikitext2 dataset~\cite{2016_arxiv_wikitext2} for its ability to reflect the performance of LLMs. We also evaluate the reasoning capabilities on MMLU and CommonSenseQA benchmarks.
We run all evaluations using NVIDIA's A100 GPU (with 80GB HBM) on our servers. We extend the code of KVquant~\cite{2024_arxiv_kvquant}, GPTVQ~\cite{2024_arxiv_gptvq}, and lm-eval-harness~\cite{2024_arxiv_lmeval} to evaluate our methods. 

We evaluate quantization error, perplexity-to-footprint trade-offs, perplexity results, reasoning accuracy, different block sizes, and perplexity-to-recycled values to demonstrate the superiority of NxFP. In addition, we cumulatively add NanoMantissa (\textbf{NM}), Adaptive Microexponents (\textbf{NM+AM}), and code recycling (\textbf{NM+AM+CR}) techniques to isolate their benefits.

\begin{table}[ht!]
    \centering
    \includegraphics[width=0.99\linewidth]{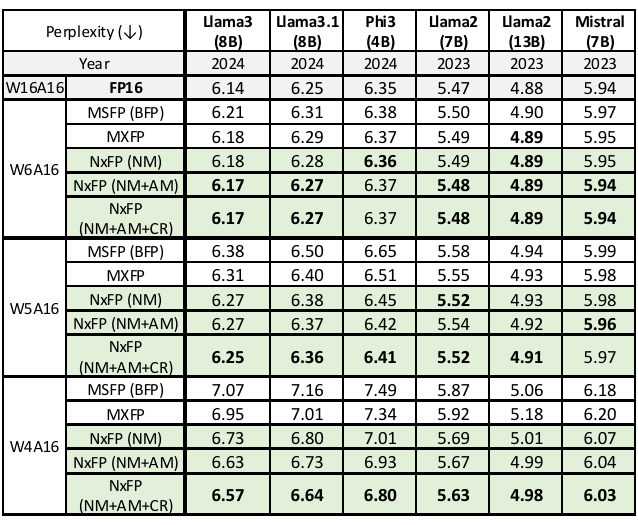}
    \caption{Weight-only quantization results on five modern LLMs, where NxFP consistently achieves the best perplexity. W4A16 denotes 4-bit weight and 16-bit activations.}
    \label{table:perplexity}
\end{table}

\subsection{Quantization Error Analysis}

Figure~\ref{figure:qerr} shows the quantization error (MSE) improvement of NxFP4 over MxFP4 using the direct-cast quantization, where we can isolate the benefits of different techniques. In brief, NxFP4 reduces the quantization error by up to 45\%. The quantization error improvement is similar across all models.

We further isolate the contribution of each technique below. First, NanoMantissa helps reduce the quantization error by up to 26\%. Then, the adaptive microexponent further reduces the quantization error by 14\%. Finally, code recycling helps reduce the quantization error by 4.7\%.

\begin{figure*}[t]
    \centering
    \includegraphics[width=0.85\linewidth]{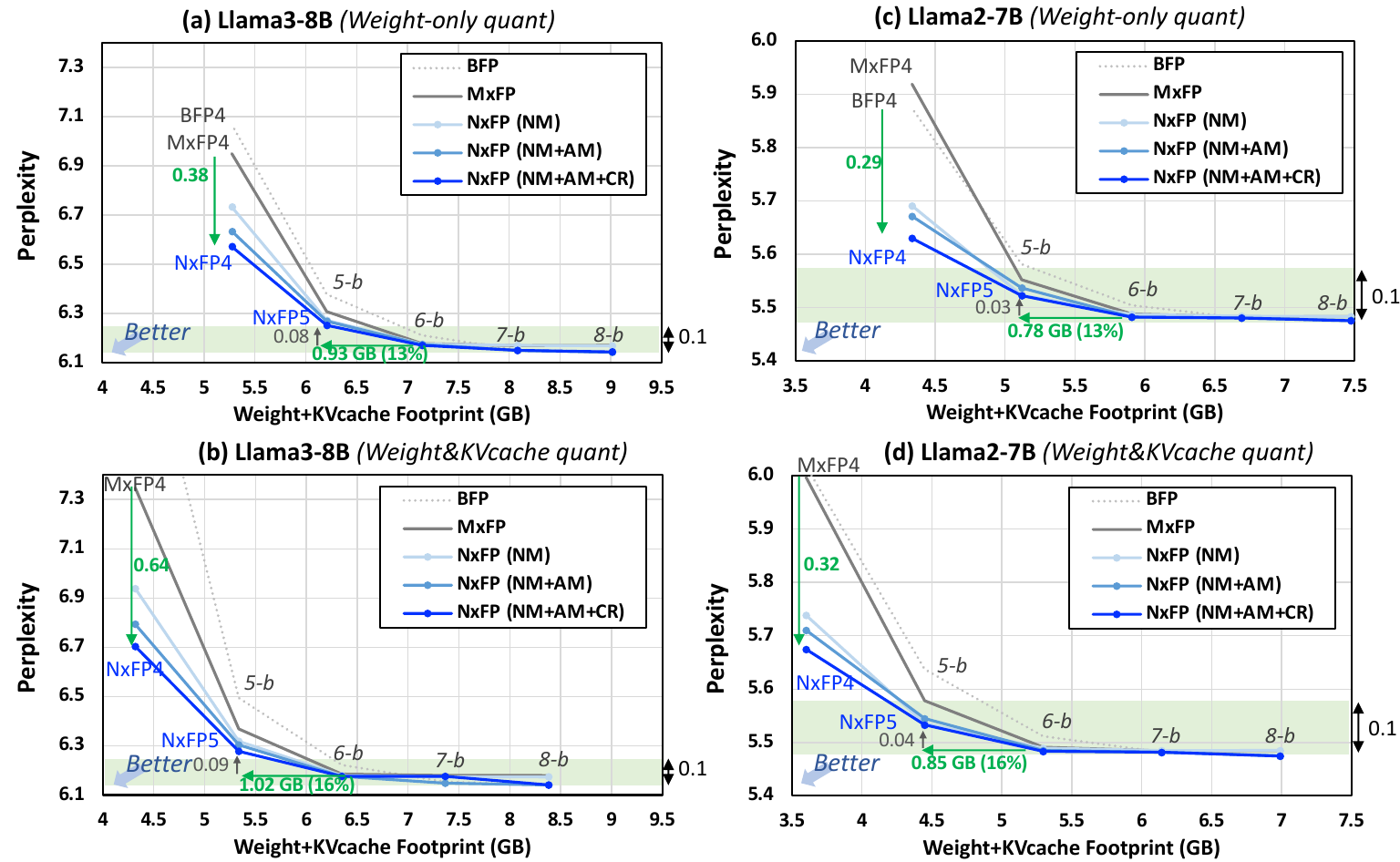}
    \caption{The perplexity-to-footprint trade-offs of weight-only quantization and weight\&KVcache quantization on (a)(b) Llama3 (8B) and (c)(d) Llama2 (7B). The NxFP consistently outperforms MxFP6 in perplexity-to-footprint trade-offs. The sequence length is set to 2K for all experiments.}
    \label{figure:ppl_footprint}
\end{figure*}

\subsection{Perplexity Degradation Analysis}

Table~\ref{table:perplexity} analyzes the perplexity degradation of weight-only quantization on modern LLMs. We mark the best perplexity with bold.
We compare MxFP against BFP and MxFP, where NxFP consistently achieves the best perplexity under 4$\sim$6 bits. We can observe that three proposed techniques successfully improves the perplexity, especially at ultra-low bits (i.e., 4-bit).

Since the trend is similar across different models, let us utilize Llama3 as the main illustrative example. The original FP16 perplexity is 6.14. Under 6-bit weight, MxFP (6.18) and BFP (6.21) degrade the perplexity by 0.4 and 0.7, respectively. On the other hand, NxFP (6.17) achieves the smallest perplexity degradation (0.03). 
Similarly, BFP5 (6.38) and MxFP5 (6.31) degrade perplexity by 0.24 and 0.17, respectively. NxFP5 (6.25), again, achieves the smallest perplexity degradation of 0.11
Finally, using 4-bit weight, the perplexity degradation of BFP, MxFP, and Nanoscaling are 0.93, 0.81, and 0.43, respectively.

\subsection{Perplexity-to-Footprint Trade-offs} \label{tradeoff}

Figure~\ref{figure:ppl_footprint} presents the perplexity-to-footprint trade-offs of Nanoscaling (NxFP), Microscaling (MxFP), and BFP on five modern LLMs. NxFP consistently achieves the best perplexity-to-footprint trade-off. We highlight the design points with $\le$0.1 perplexity degradation compared to FP16 in green. Since we perform experiments on weight-only quantization and quantizing the weight and KV cache, the x-axis showcases the total footprint of both weights and KV cache.

Figure~\ref{figure:ppl_footprint}(a) compares the perplexity-to-footprint Pareto frontier of NxFP with standard MxFP and BFP on Llama3. NxFP consistently improves the perplexity using a smaller model footprint. 
First, NxFP4 can significantly reduce the perplexity by 0.38 compared to MxFP4.
Second, NxFP5 reduces 0.93 GB (13\%) memory footprint with only 0.08 perplexity degradation compared to MxFP6. 
Third, NxFP5 can satisfy the $\le$0.1 perplexity degradation requirement and reduces 10GB footprint compared to FP16. 

Figure~\ref{figure:ppl_footprint}(b) presents the trade-offs when we quantize both the weights and the KV cach. 
First, NxFP4 can significantly reduce the perplexity by 0.64 compared to MxFP4.
Further, NxFP5 reduces 1.02 GB (16\%) memory footprint with only 0.09 perplexity degradation compared to MxFP6, which nearly satisfies the $\le$0.1 perplexity degradation requirement compared to FP16.

A similar trend can be observed on Llama2 in Figure~\ref{figure:ppl_footprint}(c) and Figure~\ref{figure:ppl_footprint}(d). Compared to MxFP6, NxFP5 can reduce the footprint by 0.78$\sim$0.85 GB. Compared to FP16, NxFP5 reduces the footprint by 8.43GB while only degrading the perplexity by 0.05.

\subsection{Accuracy Degradation Analysis on MMLU Reasoning}

\begin{figure}[t]
    \centering
    \includegraphics[width=0.99\linewidth]{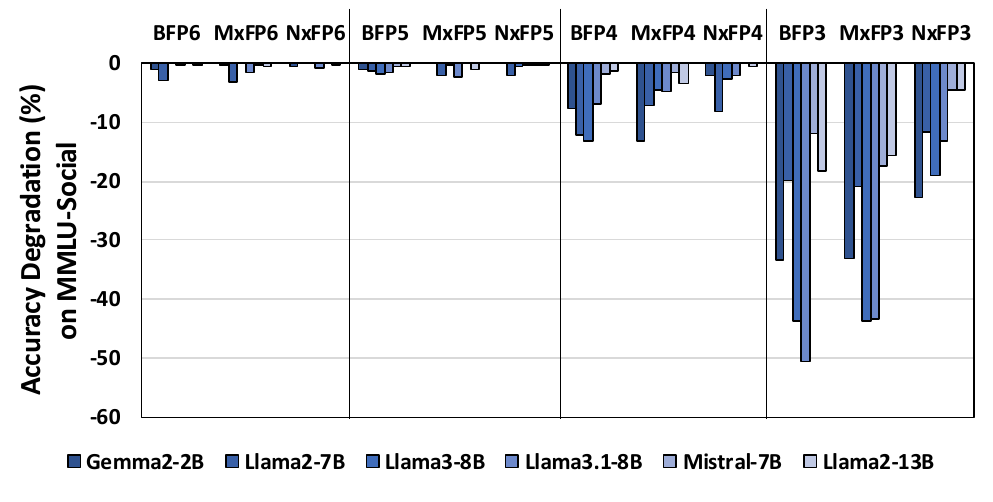}
    \caption{Accuracy degradation analysis of different quantization techniques on MMLU (Social Sciences).}
    \label{figure:acc_mmlu}
\end{figure}

Figure~\ref{figure:acc_mmlu} summarizes the accuracy degradation of various LLMs (Llama3, Llama3.1, Gemma2, Llama2, and Mistral) on MMLU-SocialScience. 

There are three takeaways.
First, NxFP significantly mitigates the accuracy degradation on low bitwidths, i.e., 4-bit and 3-bit. Specifically, NxFP improves the accuracy by up to 30.2\% on MMLU-SocialScience compared to MxFP and BFP.
Second, the smaller language models (e.g., Gemma2-2B) are generally less quantizable than larger language models (Llama2-13B).
Third, Mistral is surprisingly the most quantizable model in the 7B category.

\begin{figure}[t]
    \centering
    \includegraphics[width=0.98\linewidth]{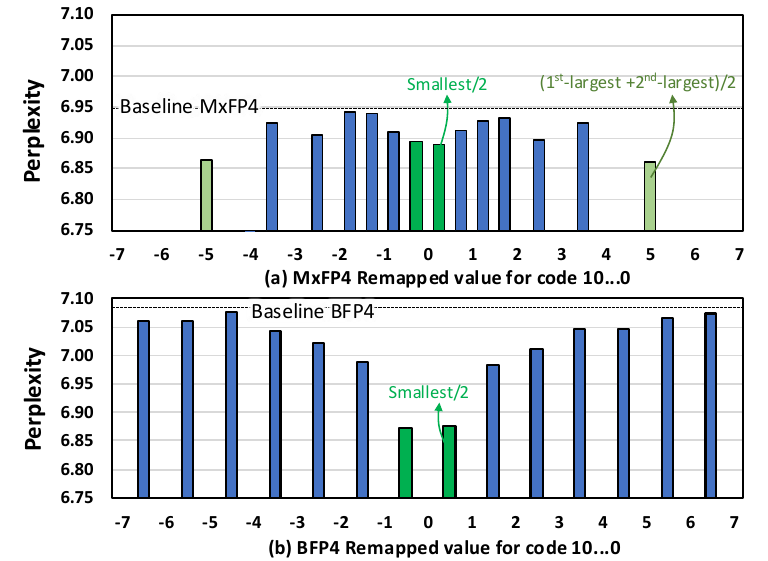}
    \caption{Perplexity of different recycled values to remap wasted code 10...0 on (a) MxFP4 and (b) BFP4 on Llama3 (8B).}
    \label{figure:sweep_recycled}
\end{figure}

\subsection{Sweeping the Remapped Value of Code Recycling}

We sweep the perplexity of different remapped value on MxFP and BFP. Although sweeping through all remapped value is possible, we consider the implementation overheads and only sweeps through the middle-points in between the original quantization levels. 

Figure~\ref{figure:sweep_recycled}(a) shows the perplexity of different remapped values on code-recycled MxFP4 for Llama3-8B. The dotted line shows the baseline perplexity of MxFP4. We show that half of the smallest quantization number and the middle-point between the 1st and 2nd-largest-quantization-value improves the perplexity the most. 
Figure~\ref{figure:sweep_recycled}(b) also shows the perplexity of different remapped values on code-recycled BFP4. The dotted line shows the baseline perplexity of BFP4. We show that half of the smallest quantization number improves the perplexity the most.

\begin{figure}[t]
    \centering
    \includegraphics[width=0.99\linewidth]{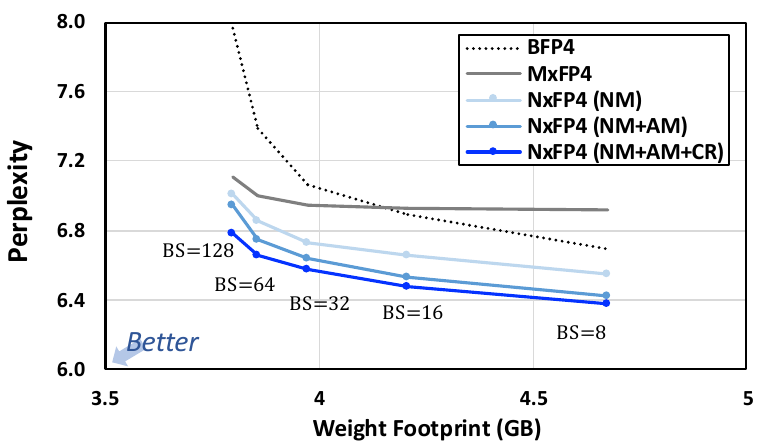}
    \caption{Perplexity-to-footprint trade-offs across different block sizes using 4 bits.}
    \label{figure:block_size}
\end{figure}

\subsection{Different Block Size}

Figure~\ref{figure:block_size} compares the perplexity-to-footprint trade-offs under different block sizes (BS) on Llama3 (8B).  
The first observation is that NxFP4 consistently outperforms MxFP4 and BFP4 regardless of the block size. 
The second observation is that MxFP4 outperforms BFP4 when the block size is large, which is because the microexponents can ensure enough element-wise dynamic range when the values are much more scattered. 

\section{Conclusions}
This paper profiles modern LLMs and identifies three challenges of low-bit Microscaling standards. In response, we propose a Nanoscaling (NxFP) format, which contains three techniques, i.e., \emph{NanoMantissa}, \emph{Adaptive Microexponent}, and \emph{Code Recycling}.

Experimental results across various modern LLMs demonstrate that our proposed methods outperform state-of-the-art MxFP by up to 0.64 in perplexity, by up to 30\% in MMLU accuracy, and 16\% smaller memory footprint. 

\section{Impact Statement}
This paper presents work whose goal is to advance the field
of Efficient Machine Learning. There are many positive societal
consequences of our work, which can make AI more accessible.

\bibliography{main}
\bibliographystyle{arxiv2025}

\end{document}